\documentstyle[amsfonts,12pt]{article} 
\title{\bf Path-Integral Aspects of Supersymmetric  Quantum Mechanics%
\thanks{Invited talk presented at the International Seminar ``Path
Integrals: Theory \& Applications'' and 5th International Conference on Path
Integrals from meV to MeV, Dubna, Russia, May 27--31, 1996.}
}
\author{\large Georg Junker
\\[3mm]
\em Institut f\"ur Theoretische Physik\\
\em Universit\"at Erlangen-N\"urnberg\\
\em Staudtstr.\ 7, D-91058 Erlangen, Germany}
\date{}
\begin{document}
\maketitle
%%%%%%%%%%%%%%%%%%%%%%%%%%%%%%%%%%%%%%%%%%%%%%%%%%%%%%%%%%%%%%%%%%%%%%%%%%%%%%%
\begin{abstract} 
In this talk we briefly review the concept of supersymmetric quantum mechanics
using a model introduced by Witten. A quasi-classical path-integral evaluation
for  this model is performed, leading to a so-called supersymmetric
quasi-classical quantization condition. Properties of this quantization
condition are compared with those derived from the standard WKB approach.
\end{abstract}
%%%%%%%%%%%%%%%%%%%%%%%%%%%%%%%%%%%%%%%%%%%%%%%%%%%%%%%%%%%%%%%%%%%%%%%%%%%%%%%
\section{Introduction}
In 1976 Nicolai \cite{Nic76} introduced supersymmetric (SUSY) quantum
mechanics as the non-relativistic version of supersymmetric quantum field
theory in
order to inves\-tigate some possible applications of SUSY to spin systems.
Independently, in 1981 Witten \cite{Wit81} considered SUSY
quantum mechanics as a toy model for studying the SUSY-breaking
mechanism in quantum field theory. During the last 15 years SUSY quantum
mechanics became an important tool in various branches of theoretical physics.
For an overview see, for example, the forthcoming monograph \cite{Jun96}.

In this lecture we will begin with a short review on the concept of SUSY in
quantum mechanics on the basis of Witten's model. Then we consider a modified
stationary-phase approach to the path-integral formulation of this model
leading to a supersymmetric version of the vanVleck-Pauli-Morette and
Gutzwiller formula for the approximate propagator and Green function,
respectively. From the poles of the approximate Green function a
quasi-classical supersymmetric quantization condition is derived and compared
with the standard WKB condition.
%%%%%%%%%%%%%%%%%%%%%%%%%%%%%%%%%%%%%%%%%%%%%%%%%%%%%%%%%%%%%%%%%%%%%%%%%%%%%%%
\section{SUSY quantum mechanics and Witten's model}
Following Nicolai \cite{Nic76} we call a quantum mechanical system
characterized by a self-adjoint Hamiltonian $H$, acting on some Hilbert space
${\cal H}$, {\em supersymmetric} if there exists a {\em supercharge}
operator $Q$ obeying the following anticommutation
relations:
\begin{equation}
\{Q,Q\}=0=\{Q^{\dagger},Q^{\dagger}\}\quad,\qquad \{Q,Q^{\dagger}\}=H\quad .
\label{SUSY}
\end{equation}
An immediate consequence of these relations is the conservation of the
supercharge and the non-negativity of the Hamiltonian,
\begin{equation}
[H,Q]=0=[H,Q^{\dagger}]\quad ,\qquad H\geq 0\quad .
\label{HQ}
\end{equation}

In 1981 Witten \cite{Wit81} introduced a simple model of supersymmetric
quantum mechanics. It is defined on the Hilbert space
${\cal H}:=L^2({\Bbb R})\otimes{\Bbb C}^2$, that is, 
it characterizes 
a spin-$\frac{1}{2}$-like particle (with mass $m>0$) moving along the
one-dimensional Euclidean line ${\Bbb R}$.
In constructing a supersymmetric Hamiltonian on
${\cal H}$ let us first introduce a bosonic operator $b$ and a fermionic
operator $f$:
\begin{equation}
\begin{array}{ll}
\displaystyle
b:L^2({\Bbb R})\to L^2({\Bbb R})\quad,\qquad &\displaystyle
b:=\frac{1}{\sqrt{2}}\Bigl(\frac{\partial}{\partial x}+
                      \Phi (x)\Bigr)\quad,\\[4mm]
\displaystyle
f:{\Bbb C}^2\to{\Bbb C}^2\quad,\qquad &\displaystyle
f:=\left(\begin{array}{@{}cc@{}}0&0\\ 1&0\end{array}\right)\quad,
\end{array}
\label{bf}
\end{equation}
where the {\em SUSY potential} $\Phi :{\Bbb R}\to{\Bbb R}$ is assumed 
to be continuously differentiable. 
Obviously these operators obey the commutation and anticommutation relations
\begin{equation}
[b,b^{\dagger}]=\Phi '(x)\quad,\qquad \{f,f^{\dagger}\}=1\quad ,
\label{bbff}
\end{equation}
and allow us to define a suitable supercharge
\begin{equation}
Q:=\frac{\hbar}{\sqrt{m}}\,b\otimes f^{\dagger}=\frac{\hbar}{\sqrt{m}}
\left(\begin{array}{@{}cc@{}}0&b\\ 0&0\end{array}\right)
\quad,\qquad
Q^{\dagger}=\frac{\hbar}{\sqrt{m}}\,b^{\dagger}\otimes f=
\frac{\hbar}{\sqrt{m}}\left(\begin{array}{@{}cc@{}}0&0\\
b^{\dagger}&0\end{array}\right)\quad ,
\label{Q}
\end{equation}
which obeys the required relations $\{Q,Q\}=0=\{Q^{\dagger},Q^{\dagger}\}$.
Note that $Q$ is a combination of a generalized bosonic annihilation operator
and a fermionic creation operator.
Finally, we may construct a supersymmetric quantum system by defining
the Hamilton\-ian in such a way that also the second relation in (\ref{SUSY})
holds,
\begin{equation}
H:=\{Q,Q^{\dagger}\}=\frac{\hbar^2}{m}
\left(\begin{array}{@{}cc@{}}bb^{\dagger}&0\\ 0&b^{\dagger}b
\end{array}\right)=
\left(\begin{array}{@{}cc@{}}H_{+}&0\\ 0&H_{-}\end{array}\right)
\quad ,
\label{H}
\end{equation}
with
\begin{equation}
H_{\pm}:=\frac{\hbar^2}{2m}
\left[-\frac{\partial^2}{\partial x^2}+\Phi ^2(x)\pm\Phi '(x)\right]\geq 0
\label{Hpm}
\end{equation}
being standard Schr\"odinger operators acting on $L^2({\Bbb R})$.

The supersymmetry (\ref{SUSY}) of a quantum system is said to be a good
symmetry (good SUSY) if the ground-state energy of $H$ vanishes.
In the other case, $\mbox{inf spec }H>0$, SUSY is said to be broken.
For good SUSY the ground state of $H$ either belongs to $H_{+}$ or $H_{-}$
and is given by
\begin{equation}
\varphi _{0}^{\pm}(x)=\varphi _{0}^{\pm}(0)\exp\left\{\pm\int\limits_{0}^{x}
{\rm d}z\,\Phi (z)\right\}\quad .
\label{varphi0}
\end{equation}
Obviously, depending on the asymptotic behavior of the SUSY potential one of
the two functions $\varphi _{0}^{\pm}$ will be normalizable (good SUSY) or
both are not normalizable (broken SUSY). To be more explicit, let us
introduce the {\em Witten index}, which (according to the Atiyah-Singer index
theorem) depends only on the asymptotic values of $\Phi $:
\begin{equation}
\Delta :=\mbox{ind }b=\mbox{dim ker }H_{-}-\mbox{dim ker }H_{+}=
\textstyle\frac{1}{2}[\mbox{sgn }\Phi (+\infty )-\mbox{sgn }\Phi (-\infty )]
\quad .
\label{Delta}
\end{equation}
Hence, for good SUSY we have $\Delta =\pm 1$ with the ground state belonging
to $H_{\mp}$. For broken SUSY we have $\Delta =0$.
Due to SUSY it is also easy to show \cite{Jun96}
that $H_{+}$ and $H_{-}$ are essentially
iso-spectral, that is, their strictly positive eigenvalues are identical.
These spectral properties of $H_{\pm}$ are summarized in the following table,
\begin{equation}
\begin{array}{@{}lll@{}}
\Delta =+1\quad &: \qquad & E_{n}^+=E_{n+1}^->0\quad,\qquad E_{0}^-=0\quad ,\\
\Delta =-1\quad &: \qquad & E_{n}^-=E_{n+1}^+>0\quad,\qquad E_{0}^+=0\quad ,\\
\Delta =0\quad  &: \qquad & E_{n}^-=E_{n}^+>0 \quad ,
\end{array}
\label{isospectral}
\end{equation}
where $E^{\pm}_{n}$, $n=0,1,2,\ldots$, denotes the ordered set of eigenvalues
of $H_{\pm}$ with $E^{\pm}_{n}<E^{\pm}_{n+1}$. For simplicity, we have assumed
purely discrete spectra.
%%%%%%%%%%%%%%%%%%%%%%%%%%%%%%%%%%%%%%%%%%%%%%%%%%%%%%%%%%%%%%%%%%%%%%%%%%%%%%
\section{Quasi-classical path-integral evaluation}
Let us now consider a quasi-classical evaluation of the
path integral associated with the propagator of the pair of Hamiltonians
(\ref{Hpm}),
\begin{equation}
K_{\pm}(b,a;\tau ):=\langle b|{\rm e}^{-({\rm i}/\hbar)\tau H_{\pm}}|a\rangle=
\int\limits_{x(0)=a}^{x(\tau )=b}{\cal D}x\,
\exp\left\{{\rm i}S_{0}[x]\mp{\rm i}\varphi [x]\right\}\quad ,
\label{K}
\end{equation}
where the so-called {\em tree action} $S_{0}$ and {\em fermionic phase}
$\varphi $ are given by
\begin{equation}
S_{0}[x]:=\frac{1}{2}\int\limits_{0}^{\tau\hbar/m}{\rm d}t
\left[\dot{x}^2(t)-\Phi ^2(x(t))\right]\quad,\qquad
\varphi [x]:=\frac{1}{2}\int\limits_{0}^{\tau\hbar/m}{\rm d}t\,\Phi '(x(t))
\quad .
\label{S0varphi}
\end{equation}
Usually, in the stationary-phase approximation one considers the quadratic
fluctuations about stationary paths of the full action
$S_{\pm}[x]:=S_{0}[x]\mp\varphi [x]$. As a modification of this approach we
have suggested \cite{InJu91} to consider the fluctuations about
the stationary paths of the tree action.
These so-called {\em quasi-classical} paths \cite{JuMa94} are denoted by
$x_{qc}$, that is, $\delta S_{0}[x_{qc}]=0$.
Hence, we approximate the tree action to second order in
$\eta (t):=x(t)-x_{qc}(t)$ and consider the fermionic phase only along the
quasi-classical path:
\begin{equation}
S_{\pm}[x]\approx S_{0}[x_{qc}]\mp\varphi [x_{qc}]+
\frac{1}{2}\int\limits_{0}^{\tau\hbar/m }{\rm d}t
\left[\dot{\eta }^2(t)+\frac{1}{2}{\Phi ^2}''(x_{qc}(t))\,\eta ^2(t)\right]
\quad .
\label{Sapprox}
\end{equation}
Performing the resulting Fresnel-type path integral we arrive at a
supersymmetric version of the vanVleck-Pauli-Morette formula \cite{InJu94},
\begin{equation}
K_{\pm}(b,a;\tau )\approx\sum_{x_{qc}}^{\tau~{\rm fixed}}
\sqrt{\frac{\rm i}{2\pi }\left|\frac{\partial^2S_{0}[x_{qc}]}
{\partial a\partial b}\right|}\,
\exp\left\{{\rm i}S_{0}[x_{qc}]-{\rm i}\frac{\pi }{2}\mu [x_{qc}]
\mp{\rm i}\varphi [x_{qc}]\right\}\quad ,
\label{Kapprox}
\end{equation}
where $\mu[x_{qc}] $ denotes the Morse index and equals the number of
conjugated points along $x_{qc}$.

In order to obtain some spectral information
we pass over from the propagator to the Green function
\begin{equation}
\begin{array}{@{}rl@{}}
G_{\pm}(b,a;\varepsilon ):=&\displaystyle
\left\langle b\left|\frac{1}{H_{\pm}-
\frac{\hbar^2}{2m}\,\varepsilon}\right|a\right\rangle\\[2mm]
=&\displaystyle
\frac{1}{{\rm i}\hbar}\int\limits_{0}^\infty {\rm d}\tau \,
K_{\pm}(b,a;\tau )\,\exp\{{\rm i}\tau \varepsilon \hbar/2m\}\quad,
\qquad\mbox{Im }\varepsilon >0\quad .
\end{array}
\label{G}
\end{equation}
A stationary-phase evaluation of this integral leads to the supersymmetric
version of Gutzwiller's formula \cite{InJu94}:
\begin{equation}
\begin{array}{@{}ll@{}}
G_{\pm}(b,a;\varepsilon )\approx &\displaystyle
\frac{m}{{\rm i}\hbar^2}\left|\Bigl(\varepsilon -\Phi ^2(a)\Bigr)
\Bigl(\varepsilon -\Phi ^2(b)\Bigr)\right|^{-1/4}\\[3mm]
&\times\displaystyle
\sum_{x_{qc}}^{\varepsilon ~{\rm fixed}}\exp\left\{{\rm i}W_{0}[x_{qc}]-{\rm
i}\frac{\pi }{2}\nu [x_{qc}]\mp{\rm i}\varphi[x_{qc}]\right\}\quad .
\end{array}
\label{Gapprox}
\end{equation}
Here $W_{0}[x_{qc}]:=\int_{x_{qc}}{\rm d}x\sqrt{\varepsilon -\Phi ^2(x)}$
denotes Hamilton's characteristic function (associated with the tree action
$S_{0}[x_{qc}]$) and $\nu [x_{qc}]$ is the Maslov index,
which equals the number of turning points along $x_{qc}$.
For a single-well shape of $\Phi ^2$ the sum in (\ref{Gapprox}) can explicitly
be performed \cite{InJu91,Jun96}.
The poles of the resulting formula give rise to the
{\em quasi-classical supersymmetric} (qc-SUSY) quantization condition
\begin{equation}
\fbox{$\displaystyle
\int\limits_{x_{L}}^{x_{R}}{\rm d}x\sqrt{\varepsilon -\Phi ^2(x)}=\pi
\left(n+\frac{1}{2}\pm\frac{\Delta }{2}\right)$}\quad ,
\label{qcSUSY}
\end{equation}
where $x_{L/R}$ denote  the left and right turning points of $x_{qc}$,
$\Phi ^2(x_{L/R})=\varepsilon=E/\frac{\hbar^2}{2m} $.
This quantization condition may be compared
with the usual WKB condition derived from a stationary-phase approximation of
the full action $S_{\pm}$,
\begin{equation}
\int\limits_{q^\pm_{L}}^{q^\pm_{R}}{\rm d}x
\sqrt{\varepsilon -\Phi ^2(x)\mp\Phi '(x)}=
\pi \left(n+\frac{1}{2}\right)\quad ,
\label{WKB}
\end{equation}
with $q^\pm_{L/R}$ as the classical turning points,
$\Phi ^2(q^{\pm}_{L/R})\mp\Phi '(q^{\pm}_{L/R})=\varepsilon=
E/\frac{\hbar^2}{2m}$.
%%%%%%%%%%%%%%%%%%%%%%%%%%%%%%%%%%%%%%%%%%%%%%%%%%%%%%%%%%%%%%%%%%%%%%%%%%%%%%
\section{Discussion}
It should be emphasized that the eigenvalues obtained from the qc-SUSY
approxi\-mation (\ref{qcSUSY}) respect all of the spectral properties given in
(\ref{isospectral}). On the contrary, this is in general not the case for the
WKB spectrum (\ref{WKB}). Furthermore, (\ref{qcSUSY}) leads to the
exact bound-state spectrum for all so-called shape-invariant potentials. These
shape-invariant potentials are known to give rise to factorizable Hamiltonians
and hence the eigenvalue problem is easily solvable via the well-known
factorization method \cite{InHu51} or an explicit path integral evaluation
\cite{InKrGe92}. This exactness can also be achieved via the WKB
approximation, which, however, requires ad hoc Langer-type modifications. Here
the question naturally arises: Why is the qc-SUSY approximation exact for
those shape-invariant potentials? One possible explanations of this fact via
the Nicolai mapping has been discussed in \cite{Jun96}.
An alternative explanation may be based on the path-integral generalization
of the Duistermaat-Heckman theorem \cite{Ati84}.

For various not exactly solvable potentials numerical investigations
\cite{InJu94} indicate that for analytical SUSY potentials
the qc-SUSY approximation always overestimates the exact energy eigenvalues.
Whereas, at least for the case of broken SUSY, the WKB approximation gives an
underestimation. For a detailed numerical investigation of this
level-ordering phenomenon see the monograph \cite{Jun96}, where also other
applications of the above quasi-classical approach are discussed.
%%%%%%%%%%%%%%%%%%%%%%%%%%%%%%%%%%%%%%%%%%%%%%%%%%%%%%%%%%%%%%%%%%%%%%%%%%%%%%
\section*{Acknowlegements}
I would like to thank Hajo Leschke and Peter van Nieuwenhuizen for drawing my
attention to the Duistermaat-Heckman theorem. I am also very grateful to
Bernhard Bodmann and Simone Warzel for clarifying discussions on the
Duistermaat-Heckman theorem, including its path-integral generalization, and
for their comments and suggestions on this manuscript.
%%%%%%%%%%%%%%%%%%%%%%%%%%%%%%%%%%%%%%%%%%%%%%%%%%%%%%%%%%%%%%%%%%%%%%%%%%%

%%%%%%%%%%%%%%%%%%%%%%%%%%%%%%%%%%%%%%%%%%%%%%%%%%%%%%%%%%%%%%%%%%%%%%%%%%%
\end{document}